\documentclass[preprint,showpacs,showkeys,aps,prb]{revtex4}
\usepackage[english]{babel}
\usepackage[dvips]{graphicx}

\begin{document}

\title{Another Hamiltonian ``Thermostat'' -- Comments on arXiv 1203.5968, 1204.4412, 1205.3478,
and 1206.0188}

\author{
Wm. G. Hoover                                   \\
Ruby Valley Research Institute                  \\
Highway Contract 60, Box 601                    \\
Ruby Valley, Nevada 89833                       \\
}

\date{\today}



\vspace{0.1cm}

\begin{abstract}
Campisi, Zhan, Talkner, and H\"anggi state, in promoting a new logarithmic computational
thermostat [ arXiv 1203.5968 and 1204.4412 ] , that (thermostated) Nos\'e-Hoover
mechanics is not Hamiltonian.  First I point out that Dettmann clearly showed
the Hamiltonian nature of Nos\'e-Hoover mechanics.  The trajectories $\{ \ q(t) \ \}$
generated by Dettmann's Hamiltonian are {\it identical} to those generated by Nos\'e-Hoover
mechanics.  I also observe that when the (Hamiltonian) Campisi thermostat is applied to
``nonequilibrium'' heat transfer problems some very interesting, and somewhat paradoxical,
phase portraits result.  See too Marc Mel\'endez' nice arXiv 1205.3478 as well as our joint
work arXiv 1206.0188.
\end{abstract}

\maketitle

In 1984 Shuichi Nos\'e discovered a deterministic, time-reversible, logarithmic
thermostat\cite{b1}.  This computational thermostat imposes a time-averaged
kinetic temperature $kT \equiv m\langle \ v^2 \ \rangle $
through a thermostated Hamiltonian.  His Hamiltonian includes a ``time-scaling variable'',
$s$, along with its conjugate momentum $p_s$.  Consider the simplest interesting example.
For a single harmonic oscillator (with mass $m$, force constant $\kappa$, relaxation time
$\tau$ and Boltzmann's constant $k$ all set equal to unity) Nos\'e's
thermostated Hamiltonian is :
$$
2{\cal H}_{\rm Nos\acute{e}} = [ \ (p/s)^2 + q^2 + p_s^2 + T\ln (s^2) \ ] \ .
$$
The equations of motion which follow from Nos\'e's Hamiltonian ,
$$
\dot q = (p/s^2) \ ; \ \dot p = -q \ ; \ \dot s = p_s \ ; \ \dot p_s = (p^2/s^3) - (T/s) \ ,
$$
are somewhat ``stiff'' because $s$ can become arbitrarily small.  ``Scaling the time'' in these
equations of motion, by a factor of $s$, gives a {\it new}, and better behaved, set :
$$
\dot q = (p/s) \ ; \ \dot p = -sq \ ; \ \dot s = sp_s \ ; \ \dot p_s = (p/s)^2 - T \ .
$$
Then, to convert to the simpler ``Nos\'e-Hoover'' form: introduce $v = (p/s)$
and $\zeta = p_s$ :
$$
\dot q = v \ ; \ \ddot q = \dot v = -q - \zeta \dot q = -q -\zeta v\ ;
 \ \dot \zeta = [ \ \dot q^2 - T \ ] = [ \ v^2 - T \ ] \ .
$$
The time scaling used to obtain the Nos\'e-Hoover equations suggests (wrongly, it turns out)
that they are  ``non-Hamiltonian''.  Though well-behaved, these Nos\'e-Hoover motion equations
 are not ergodic.  For the harmonic oscillator they  have a wide variety of periodic,
nearly-periodic, and chaotic, solutions\cite{b2}.

On the other hand, in July 1996, Carl Dettmann discovered that a {\it different}, but closely
related, Hamiltonian ,
$$
2{\cal H}_{\rm Dettmann} \equiv s{\cal H}_{\rm Nos\acute{e}} =
(p^2/s) + s[ \ q^2 + p_s^2 + T\ln (s^2)\ ] \equiv 0 \ ,
$$
gives directly the Nos\'e-Hoover motion equations but {\it without any time scaling}\cite{b3,b4,b5}.
The equations of motion from Dettmann's Hamiltonian are as follows :
$$
\dot q = (p/s) \ ; \ \dot p = - sq \ ; \ \dot s = sp_s \ ; \
\dot p_s = (1/2)(p/s)^2 - [ \ (1/2)q^2 + (1/2)p_s^2 + T\ln s \ ] - T \ .
$$
Because Dettmann's Hamiltonian is identically equal to zero, the combination in square
brackets is equal to $-(1/2)(p/s)^2$ :
$$
[ \ (1/2)q^2 + (1/2)(p_s^2 + T\ln s \ ] \equiv -(1/2)(p/s)^2 \ .
$$  
Introducing $v = (p/s)$ and $\zeta = p_s$ {\it again} produces the Nos\'e-Hoover equations,
but this time {\it without the need for any time scaling} :
$$
\dot q = v \ ; \ \ddot q = \dot v = -q - \zeta \dot q = -q -\zeta v\ ;
 \ \dot \zeta = [ \ \dot q^2 - T \ ] = [ \ v^2 - T \ ] \ .
$$
 
A year later Dettmann and Gary Morriss published yet {\it another} Hamiltonian form\cite{b4} :
$$
2{\cal H}_{\rm DM} \equiv e^{-Q}[ \ p^2 + P^2 \ ] +e^{+Q}[ \ q^2 + 2TQ \ ] \equiv 0 \ .
$$
This Dettmann-Morriss Hamiltonian, though slightly different, leads, in a very similar way,
to the same Nos\'e-Hoover motion equations. This approach was soon rediscovered by Bond,
Leimkuhler, and Laird\cite{b6}, who refer to Dettmann's original Hamiltonian as the
Nos\'e-Poincar\'e Hamiltonian. All these discoveries make the Hamiltonian nature of the
Nos\'e-Hoover equations quite clear.

Campisi {\it et alii}\cite{b7} cited Klages' book\cite{b8} as the source of their ``nonHamiltonian''
characterization of Nos\'e-Hoover mechanics.  In that book Klages warns the reader: ``one should
carefully distinguish between traditional Hamiltonian formulations of classical mechanics and
the generalized Hamiltonian formalism outlined [ in the book ].''  It seems to me that one can
take this warning too seriously. In several helpful and stimulating private
communications Campisi, H\"anggi, and Klages emphasized the arbitrary value of Dettmann's
Hamiltonian, zero, as well as the failure of the $(q,v,\zeta)$ Nos\'e-Hoover equations to
follow Liouville's Theorem.

Admittedly, Dettmann's zeroing of the Hamiltonian could be accomplished in several ways.  One
way is to add a term proportional to $s$ , chosen in conjunction with the initial conditions
in order to make the Hamiltonian vanish.  This is essentially the ``Poincar\'e'' transformation
noticed by Bond, Leimkuhler, and Laird.  Continuing with the simple harmonic oscillator example,
with $T$ additionally chosen equal to unity :
$$
\{ \ q,p,s,p_s \ \} = \{ \ 1,0,1,1 \ \} \ \ {\rm [ \ initial \ conditions \ ]} \ ,
$$
the Hamiltonian becomes :
$$
{\cal H}_{\rm Another} \equiv (s/2)[ \ (p/s)^2 + q^2 + p_s^2 \ ] + s\ln (s) - s \ .
$$
The corresponding equations of motion that follow from it (and satisfy Liouville's theorem)  are :
$$                                                                                                                      
\{ \ \dot q = (p/s) \ ; \ \dot p = - sq \ ; \ \dot s = sp_s \ ; \                                                       
\dot p_s = (1/2)[ \ (p/s)^2 - q^2 - p_s^2 \ ] - \ln s  \ \} \ .                                                         
$$
The solution of these equations is identical to that from the Nos\'e-Hoover equations,
$$                                                                                                                      
\{ \ \dot q = v \ ; \ \dot v = -q - \zeta v \ ; \ \dot \zeta = v^2 - 1 \ \} \ ,                                                  
$$
where $v = (p/s)$, $\zeta = p_s$ , and the initial values are $\{ \ q,v,\zeta \ \} = \{ \ 1,0,1 \ \} \ $.

There is no obvious way to introduce a second temperature into the Nos\'e, Dettmann, or
Dettmann-Morriss Hamiltonian.  Nevertheless multi-temperature problems can be treated easily by generalizing
the Nos\'e-Hoover equations of motion to control sets of velocities through a set of
friction coefficients\cite{b5}.  By sandwiching Newtonian degrees of freedom between two sets of
boundary particles (a ``hot'' set and a ``cold'' set) it is easy to simulate steady heat flow.  It
is well-established that such dissipative systems give multifractal strange attractors in
the full system+thermostats phase space\cite{b5}.

Campisi {\it et alii} claim that their own Hamiltonian ,
$$
{\cal H}_{\rm Campisi} = {\cal H}_{\rm usual} + (kT/2)\ln (\delta^2 + S^2) + (P^2/2M) \ ,
$$
plus an unspecified (and crucial) weak coupling between the system variables $\{ \ q \ \}$ and the
thermostat variable $S$ is an improvement\cite{b7}.  If their thermostat is more easily matched in
laboratory experiments then it is indeed a step forward.  But, if a multi-temperature Campisi
Hamiltonian, including $\Sigma [ \ (kT_i/2)\ln (\delta^2 + S_i^2) \ ]$, {\it could} impose more
than a single temperature on selected degrees of freedom, either Liouville's theorem or the
fractal phase-space structures that arise away from equilibrium {\it would} be casualties.
Typically, Hamiltonians don't give fractals.

\newpage

\section{Addendum of 30 April 2012}

\begin{figure}
\includegraphics[width=2in,angle= -90]{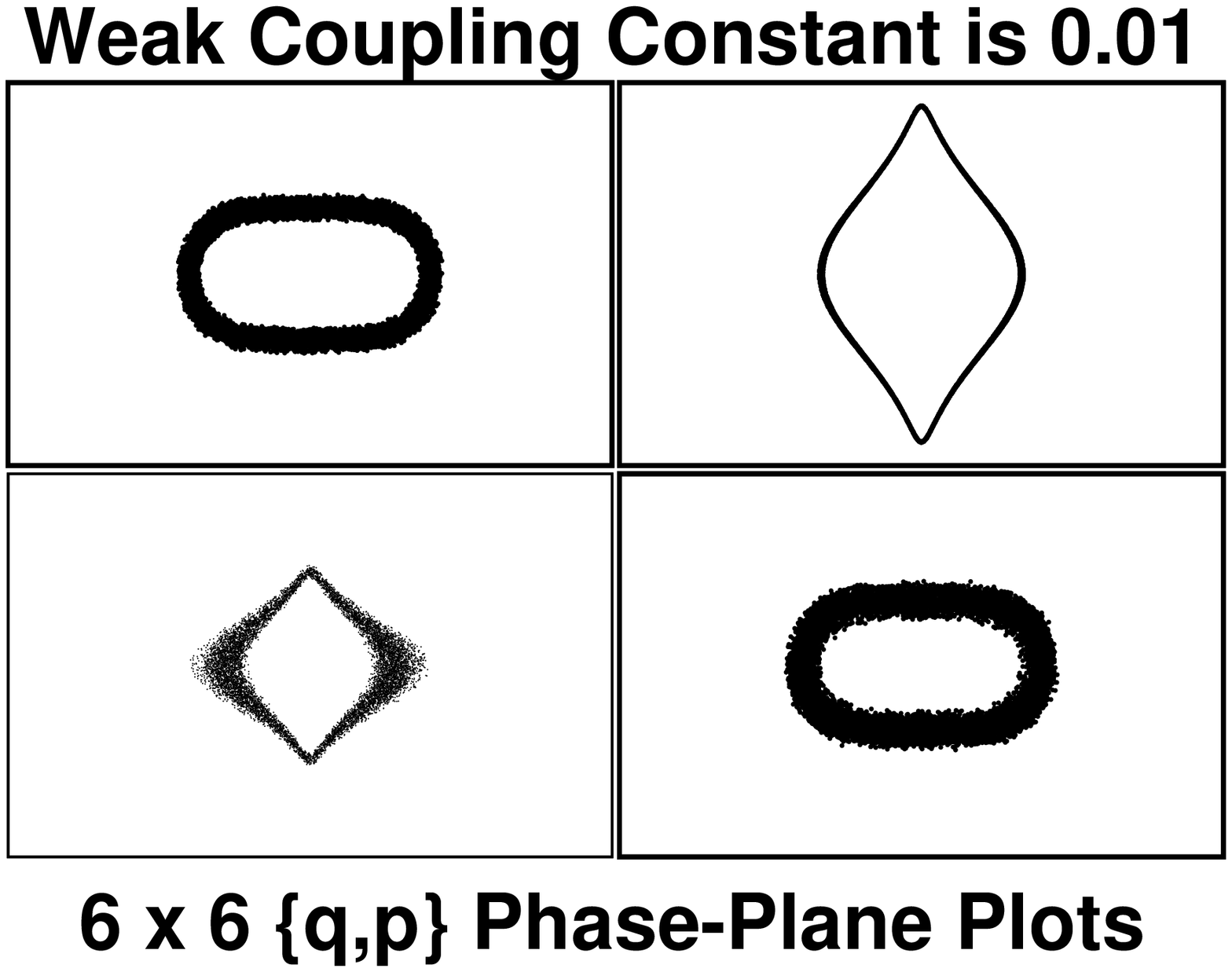}
\caption{
Phase-plane portraits with $\epsilon = 0.01$ .  The specified boundary temperatures are
0.5 and 1.5 .  Actual kinetic temperatures are $\{ \ 0.46, 0.70, 0.74, 1.40 \ \}$, left to
right, starting at the bottom.
}
\end{figure}

How can a strictly Hamiltonian thermostat satisfy Liouville's Theorem (incompressible flow
in phase space) while simultaneously giving the multifractal structures (on all scales)
associated with dissipative systems?  These two demands are contradictory, and so merit investigation.
After trying several inconclusive modifications of the simple
thermostated harmonic oscillator problem I introduced more complexity, considering a short
(two-particle) ``$\phi^4$'' chain, with one particle coupled to a ``hot'' Campisi Thermostat
and the other to a ``cold'' one.  The Figures show phase portraits for both the chain and the
thermostat particles, for the two choices $\{ \epsilon \} = \{ \ 0.01,0.10 \ \}$, using the
Hamiltonian :
$$                                                                                                                      
2{\cal H} \equiv \sum_{i=1}^2 [ \ T_i\ln(S_i^2 + \epsilon) + P_i^2 \ ] +                                                      
\epsilon [ \ (S_1-x_1)^2 + (S_2-x_2)^2 \ ] + (x_1^4 + x_2^4)/2 + (x_1-x_2)^2 +                                          
p_1^2 + p_2^2 \ .                                                                                                       
$$
The particle phase-plane $\{ \ q,p \ \}$ trajectories are shown in Figures 1 and 2.  The two
imposed temperatures, nominally 1.5 and 0.5, and the weak-coupling parameter $\epsilon = 0.01$
and $0.10$ , provide typical, but quite surprising, results.

The nature of these solutions is at the least ``odd'', from the standpoint of the Second Law
of Thermodynamics.  They provide a large temperature gradient, but with nowhere for the heat
to flow!  There are no energy sources or sinks in these problems.  Evidently these Hamiltonian
systems cannot be dissipative, {\it despite} their temperature gradients, and have instead
perfectly well-behaved solutions satisfying Liouville's incompressible
Theorem, $(df/dt) \equiv 0$ .

This odd behavior deserves a more thorough investigation than I can undertake here.  But the broad
outlines seem clear.  The nature of the Hamiltonian thermostat coupling is crucial.  If the coupling
is weak then the thermostats oscillate independently of the ``system'' (here a two-particle chain).  If the
coupling is strong then the thermostats no longer impose the desired kinetic temperatures on the hot and cold
parts of the system.  These problems deserve, and will no doubt receive, detailed investigation.

\begin{figure}
\includegraphics[width=2in,angle= -90]{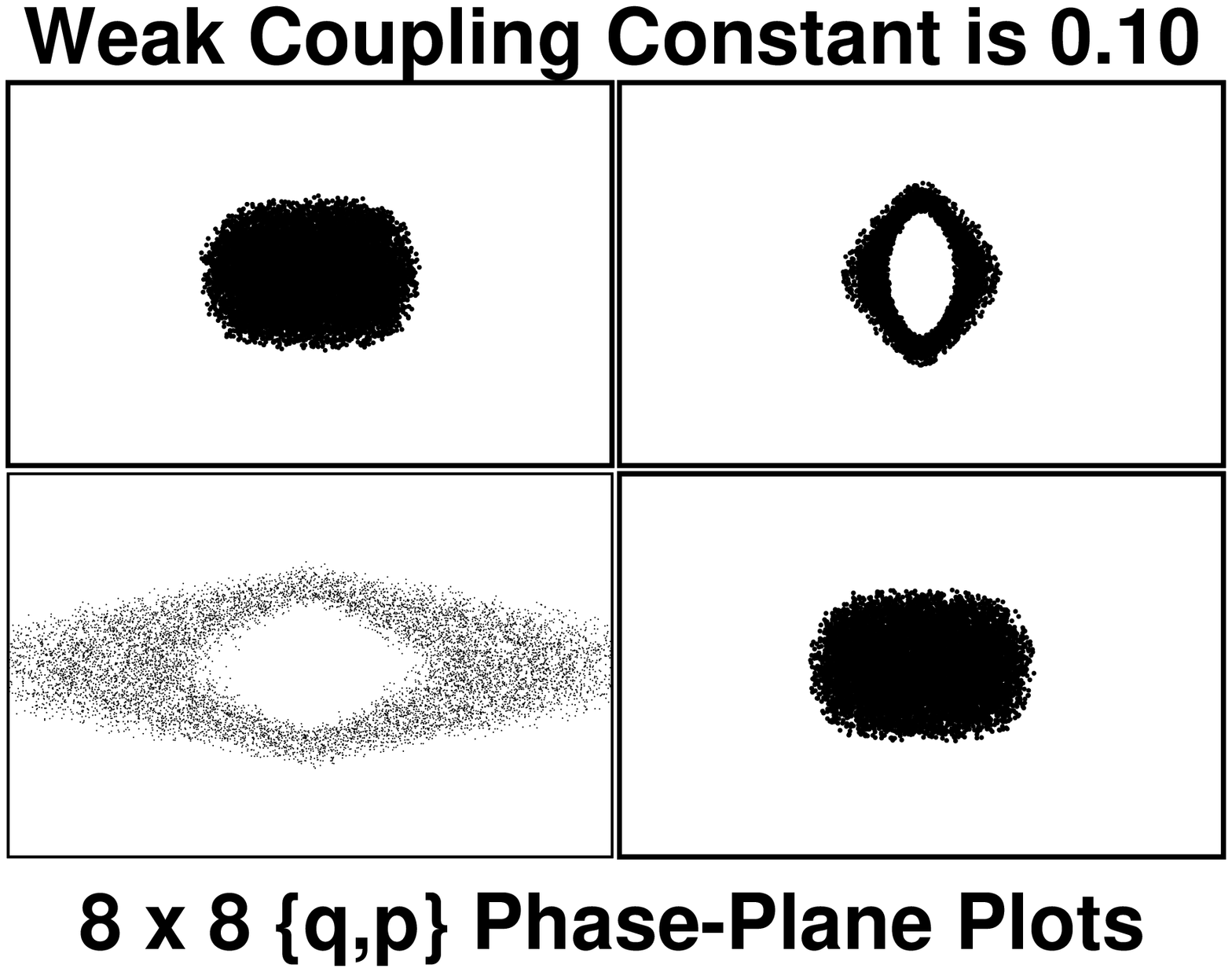}
\caption{
  Phase-plane portraits with $\epsilon = 0.10$ .  The specified boundary temperatures are
  0.5 and 1.5 .  Actual kinetic temperatures are $\{ \ 0.71, 0.53, 0.60, 0.97 \ \}$, left to
  right, starting at the bottom.
  }
\end{figure}

\newpage

\section{Addendum of 23 May 2012}

There are two other examples of non-dissipative Hamiltonian thermostats in ``Hamiltonian Dynamics of
Thermostated Systems: Two-Temperature Heat-Conducting $\phi^4$ Chains'', published in the Journal of
Chemical Physics {\bf 126}, 164113 (2007).  The ``Hoover-Leete Hamiltonian thermostat'' keeps the kinetic
energy constant through a straightforward application of Lagrangian constraints.  The ``Landau-Lifshitz
Hamiltonian thermostat'' keeps the {\it configurational} temperature constant.  {\it Both these latter
thermostats are unable to stimulate heat flow.}  The log-thermostat of Campisi {\it et alii} [
Reference 7, which will soon appear in Physical Review Letters ] is certainly the simplest example of
this anomalous behavior.

I would particularly like to thank Michele Campisi, Peter H\"anggi, Rainer Klages, and Marc Mel\'endez
for a series of stimulating and educational emails on this subject.  See also Marc's very recent and
perceptive [ ar$\chi$iv 1205.3478 ] contribution of 15 May 2012: ``On the Logarithmic Oscillator as a
Thermostat''.

\section{Addendum of 16 June 2012}

More details of the log-thermostats' shortcomings will appear in Marc Mel\'endez' and my joint work,
arXiv 1206.0188 .  Another example of Hamiltonian bases for ``nonHamiltonian'' thermostated
equations of motion appears in Wm. G. Hoover, B. Moran, C. G. Hoover, and W. J. Evans' ``Irreversibility
in the Galton Board {\it via} Conservative Classical and Quantum Hamiltonian and Gaussian Dynamics'',
Physics Letters A {\bf 133}, 114-120 (1988) .


\begin{thebibliography}{99}

\bibitem{b1} S. Nos\'e, ``Constant Temperature Molecular Dynamics Methods'',  Progress of
 Theoretical Physics Supplement {\bf 103}, 1-46, 1991.

\bibitem{b2} H. A. Posch, Wm. G. Hoover, and F. J. Vesely, ``Canonical Dynamics of the Nos\'e
Oscillator: Stability, Order, and Chaos'', Physical Review A {\bf 33}, 4253-4265 (1986).

\bibitem{b3} Wm. G. Hoover, ``M\'ecanique de Non\'equilibre \`a la Califomienne'', Physica A {\bf 240}, 1-11 (1997).

\bibitem{b4} C. P. Dettmann and G. P. Morriss, ``Hamiltonian Reformulation and Pairing of Lyapunov
Exponents for Nos\'e-Hoover Dynamics'', Physical Review E {\bf 55}, 3693-3696 (1997).


\bibitem{b5}  Wm. G. Hoover and Carol G. Hoover, {\em Time Reversibility, Computer Simulation,                          
Algorithms, and Chaos} (World Scientific, Singapore, 2012) .

\bibitem{b6} S. D. Bond, B. J. Leimkuhler, and B. B. Laird, ``The Nos\'e-Poincar\'e Method for Constant
Temperature Molecular Dynamics'', Journal of Computational Physics {\bf 151}, 114-134 (1999).

\bibitem{b7}  M. Campisi, F. Zhan, P. Talkner, and P. H\"anggi, ``Logarithmic Oscillators:                                                                      Ideal Hamiltonian Thermostats'', ar$\chi$iv 1203.5968  and 1204.4412 (2012) .

\bibitem{b8} R. Klages, {\it Microscopic Chaos, Fractals and Transport in Nonequilibrium Statistical                     
Mechanics}, (World Scientific, Singapore, 2007).

\end{thebibliography}
\end{document}